\documentclass[sigconf]{acmart}

\renewcommand\footnotetextcopyrightpermission[1]{}
\usepackage{booktabs} % For formal tables
\usepackage{flexisym}
\usepackage{bbm}
\usepackage{balance} 
\usepackage{multirow}
\usepackage{float}
\usepackage{amsmath}
\usepackage{todonotes}

\usepackage{subcaption}

\copyrightyear{2019}
\acmYear{2019}
\setcopyright{acmcopyright}
\acmConference{RecSys'19}{}{September 16--20, 2019, Copenhagen, Denmark} 
\acmPrice{15.00} 
\acmDOI{http******}
\acmISBN{ISBN *************}

\begin{document}
%\fancyhead{}
%\settopmatter{printacmref=true, printfolios=true}
\fancyhead{}

%\title{Improving Long-tail %Recommendation Over Time}
\title{Reducing Popularity Bias in Recommendation Over Time}
% Reducing Popularity Bias: An Incremental Adaptation Approach
% If Not Today, Perhaps Tomorrow: Reducing Recommendation Popularity Bias Over Time
% Not Recommended Today, Come Back Tomorrow: Reducing Popularity Bias Over Time
% 

\author{Himan Abdollahpouri}
\affiliation{%
  \institution{University of Colorado Boulder}
  \country{USA}
}
\email{himan.abdollahpouri@colorado.edu}
\author{Robin Burke}
\affiliation{%
  \institution{University of Colorado Boulder}
  \country{USA} 
}
\email{robin.burke@colorado.edu}
\begin{abstract}
Many recommendation algorithms suffer from popularity bias: a small number of popular items being recommended too frequently, while other items get insufficient exposure. Research in this area so far has concentrated on a one-shot representation of this bias, and on algorithms to improve the diversity of individual recommendation lists. In this work, we take a time-sensitive view of popularity bias, in which the algorithm assesses its long-tail coverage at regular intervals, and compensates in the present moment for omissions in the past. In particular, we present a temporal version of the well-known \textit{xQuAD} diversification algorithm adapted for long-tail recommendation. Experimental results on two public datasets show that our method is more effective in terms of the long-tail coverage and accuracy tradeoff compared to some other existing approaches.
\end{abstract}


\maketitle

\section{Introduction}

Recommender systems have an important role in e-commerce and information sites, helping users find new items. One obstacle to the effectiveness of recommenders is the problem of popularity bias \cite{bellogin2017statistical,himanaies}: collaborative filtering recommenders typically emphasize popular items (those with more ratings) over other ``long-tail'' items~\cite{longtailrecsys} that may only be popular among small groups of users. Although popular items are often good recommendations, they are also likely to be well-known. So delivering only popular items will not enhance new item discovery and will ignore the interests of users with niche tastes. It also may be unfair to the producers of less popular or newer items since they are rated by fewer users.

Most of the research addressing popularity bias has concentrated on a one-shot representation of this bias, and improving the diversity of individual recommendation lists. In other words, researchers have developed algorithms that change the recommendation lists for each user without any knowledge about how the recommender system has performed for other users. Although such approaches improve the long-tail recommendation to some extent, they miss an important opportunity for long-tail promotion: the ability of the system to compensate for previous omissions by adjusting its future output.

\begin{figure}[h]
    \centering
    \includegraphics[width=3.3in]{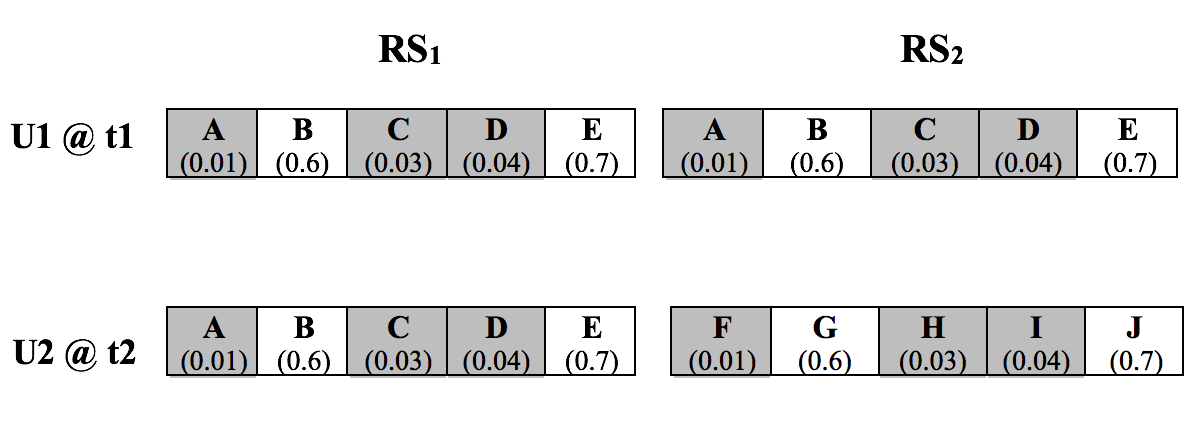}
    \vspace{-0.5cm}
    \caption{Lists of recommended items given to different users at different times by different algorithms. The grey background indicates long-tail items.}
    \label{fig:motivation}
\end{figure}

Figure ~\ref{fig:motivation} shows lists generated by two different long-tail-promoting recommendation algorithms \textit{$RS_1$} and \textit{$RS_2$} for two different users \textit{U1} and \textit{U2} arriving at two different times \textit{t1} and \textit{t2}. Popular items have a white background; long-tail items, a grey background. The outputs are superficially similar: each user gets two popular items and three long-tail items, and the averages over the popularity values of the items (shown in parentheses) are the same. However, the same three long-tail items are repeatedly generated by \textit{$RS_1$} and a greater range of such items is produced by \textit{$RS_2$}. When we are measuring long-tail coverage, the number of items and their average popularity in each list is not as important as an aggregate measure of how many different items are shown across all users. This distinction cannot be grasped looking only at individual recommendation lists in isolation; the whole recommendation set must be evaluated. In this paper, we capture this distinction through a modified version of the search diversification algorithm \textit{xQuAD} \cite{santos2010exploiting} which improves long-tail recommendation over time by adapting based on its prior performance.

%two recommendation lists for two users generated by two different recommender systems at two different times $t_1$ and $t_2$. Assume at time $t_1$ the recommender systems have generated a list of recommendations for user 1. Now, the question is what should be recommended to user 2 at time $t_2$? What can be seen is that after time $t_2$, both users have received three long tail items and two popular items. However, recommender system 1 has recommended the same items to both users while recommender system 2 has covered more unique items.  

\section{Approach}
Result diversification has been studied in the context of information retrieval, especially for web search engines, which have a related goal to find a ranking of documents that together provide a complete
coverage of the aspects underlying a query \cite{santos2015search}. EXplicit Query
Aspect Diversification (\textit{xQuAD}) \cite{santos2010exploiting} accounts for the various aspects associated with an under-specified query. Items are selected iteratively by estimating how well a given document satisfies an uncovered aspect. 

In adapting this approach, we seek to recognize the difference among users in their interest in long-tail items. Uniformly increasing the diversity of items in recommendation lists may work poorly for some users. We propose a variant that adds a personalization factor to the scoring function, based on each user's historical interest in long-tail items.

We build on the \textit{xQuAD} model to control popularity bias in recommendation algorithms over time. We assume that for a given user $u$ from the user set $U$, a ranked recommendation list $R$ of items from the item set $V$ has already been generated by a base recommendation algorithm. The task of the modified \textit{xQuAD} method is to produce a new re-ranked list $S$ ($|S|<|R|$) that manages popularity bias while still being accurate. 

In the approach for improving long-tail recommendation introduced in \cite{flairs2019} which is also based on \textit{xQuAD}, the new list is built iteratively according to the score $ s = P(v|u)+\lambda P(v,S'|u)$ where $P(v|u)$ is the likelihood of user $u \in U$ being interested in
item $v \in V$, independent of the items on the list so far, as predicted by the base recommender. The second term $P(v, S'|u)$ denotes the likelihood of user \textit{u} being interested in an item $v$ if the category of that item (long-tail vs short-head) is not in the current recommendation list $S$. 

In this work, however, since we want to take a historical view of long-tail recommendation, we do not look at $S$ to decide if an item category has already been covered or not. We, instead, look at the entire recommendation history up to the current time, $T_i$. We denote that list by $C_{T_0:T_i}$ which contains the list of items recommended from $T_0$ to $T_i$. So, we replace $P(v, S'|u)$ with $P(v, C_{T_0:T_i}'|u)$.

Intuitively, the first term in the xQuAD equation
incorporates ranking accuracy while the second term promotes
diversity between different categories of items (here, short head and long tail). The parameter $\lambda$ controls how strongly popularity bias is weighted in general. The item with the highest $s$ value is added to the output list $S$ and the process is repeated until $S$ has achieved the desired length.

The marginal likelihood $P(v, C_{T_0:T_i}'|u)$ over both
item categories long-tail ($\Gamma$) and short head ($\Gamma'$) can be computed by:

\begin{equation}\label{eq:1}
    P(v,C_{T_0:T_i}'|u)=\sum_{d \in \{\Gamma , \Gamma '\}}P(v,C_{T_0:T_i}'|d) P(d|u)
\end{equation}

Following \cite{santos2010exploiting}, we assume that the remaining items are independent of the current contents of $S$ and that the items
are independent of each other, given the short-head and long-tail categories. Under these assumptions, we can compute $P(v,C_{T_0:T_i}'|d)$ in Eq.\ref{eq:1} as

%\begin{equation} 
\begin{align} 
    P(v,C_{T_0:T_i}'|d)&= P(v|d)P(C_{T_0:T_i}'|d) \nonumber \\
    &= P(v|d)\prod_{i \in C_{T_0:T_i}} (1-P(i|d,C_{T_0:T_i})) \label{eq:2}
\end{align}
%\end{equation}

By substituting equation \ref{eq:2} into equation \ref{eq:1}, we can obtain
\begin{equation} 
     P(v,C_{T_0:T_i}'|u) =
     \sum_{d \in \{\Gamma , \Gamma '\}}P(d|u)P(v|d) \prod_{i \in C_{T_0:T_i}}(1-P(i|d,C_{T_0:T_i})) \label{eq:3}
\end{equation}
where $P(v|d)$ is equal to 1 if $v \in d$ and 0 otherwise. 

The final scoring function, therefore, is as follows:

\begin{equation} 
     s=P(u|v) +\lambda \bigg(\sum_{d \in \{\Gamma , \Gamma '\}}P(d|u)P(v|d) \prod_{i \in C_{T_0:T_i}}(1-P(i|d,C_{T_0:T_i}))\bigg) \label{eq:4}
\end{equation}
We measure $P(i|d,C_{T_0:T_i})$ in two different ways to produce two different algorithms:

\begin{itemize}
    \item \textbf{Binary}: Use the same function as $P(v|d)$, an indicator function equal to 1 when item $i$ in list $C_{T_0:T_i}$ already covers category $d$ and 0 otherwise. We call this method \textit{Time Binary} xQuAD: this is how this value is calculated using the $S$ list in the original xQuAD algorithm.
    
    \item \textbf{Time Smooth}: Another method that we introduce in this paper is to compute the fraction of category $d$ items included in the list $C_{T_0:T_i}$. We call the method that measures the $P(i|d,C_{T_0:T_i})$ in this way \textit{Time Smooth} xQuAD.

\end{itemize}
The likelihood $P(d|u)$ is the measure of user preference over different item categories. In other words, it measures how much each user is interested in short-head items versus long-tail items. We calculate this likelihood by the fraction of items in the user profile which belong to category $d$. 

In order to add the next item to $S$, we compute a re-ranking score for each item in $R \backslash S$ according to Eq. \ref{eq:4}. For an item $v' \in d$, if $C_{T_0:T_i}$ does not cover $d$, then an additional positive term will be added to the estimated user preference $P(v'|u)$, increasing the item's chance of selection and balancing accuracy and popularity bias. 

In Binary xQuAD, the product term $\prod_{i \in S}(1-P(i|d,C_{T_0:T_i}))$ is only equal to 1 if the current items
in $C_{T_0:T_i}$ have not covered the category $d$ yet. Binary \textit{xQuAD} is, therefore, optimizing for a \textit{minimal} re-ranking of the original list by including the best long-tail item it can, but not seeking diversity beyond that.

\section{Datasets and Preparation}
We tested our proposed algorithm on two public datasets. The first is the well-known MovieLens 1M dataset~\cite{movielens}. The second is the Epinions dataset, which is gathered from a consumer opinion site where users can review items \cite{massa2007trust}. Following the data reduction procedure in \cite{abdollahpouri2017controlling}, we removed users who had fewer than 20 ratings from the Epinion dataset (MovieLens already has this characteristic). We also removed distant long-tail items from each dataset using a limit of 20 ratings, a number 20 is chosen to be consistent with the cut-off for users.

After filtering, the MovieLens dataset has 6,040 users who rated 3043 movies with a total of 995,492 ratings, a reduction of about 0.4\%. Applying the same criteria to the Epinions dataset decreases the data to 220,117 ratings given by 8,144 users to 5,195 items, a reduction of around 66\%. We split the items in both datasets into two categories: long-tail ($\Gamma$) and short head ($\Gamma$') such that short-head items make up 80\% of the ratings while long-tail items have the rest. We plan to consider other divisions of the popularity distribution in future work. For MovieLens, the short-head items were those with more than 506 ratings. In Epinions, a short-head item needed only to have more than 73 ratings.

Our temporal \textit{xQuAD} algorithms operate over a series of different time epochs. To evaluate these algorithms, we split the test set into $N$ epochs -- 50 in this experiment. Investigating the effect of the number of epochs is in our plan for future work. Note that we do not need to split the data based on real time stamps, because we are not trying to learn the time-sensitive properties of users' preferences. Rather, we are only interested in simulating a succession of epochs over which the algorithm can adjust. We choose $|U_{test}|/N$ random users for each epoch where $|U_{test}|$ is the total number of users in test set.

\section{Evaluation}
The experiments compare six algorithms. Since we are concerned with ranking performance, we chose as our baseline algorithm RankALS, a pair-wise learning-to-rank algorithm. We also include the regularized long-tail diversification algorithm from \cite{abdollahpouri2017controlling} (indicated as \textit{Reg} in the figures) and two other non-temporal re-ranking approaches for long-tail recommendation (Binary xQuAD indicated as \textit{Binary} and Smooth xQuAD indicated as \textit{Smooth} in the figures) from \cite{flairs2019} against which to compare our work. The temporal versions of Binary xQuAD and Smooth xQuAD described above are labeled as  \textit{Time Binary} and \textit{Time Smooth} in the figures. We used the output from RankALS as input for the four re-ranking variants described above. We compute lists of length 100 from RankALS and pass these to the re-ranking algorithms to compute the final list of 10 recommendations for each user.\footnote{We used the implementation of RankALS in LibRec 2.0 (www.librec.net) for all experiments.}

In order to evaluate the effectiveness of algorithms in mitigating popularity bias we use four different metrics:

\textbf{Average Recommendation Popularity (ARP)}: This measure from \cite{yin2012challenging}, which calculates the average popularity of the recommended items in each list. 

\textbf{Long-tail Coverage Ratio (LCR)}: This metric measures the ratio of covered long-tail items out of all long-tail items

\begin{equation}
  LCR=\frac{\left\vert\bigcup\limits_{u \in U_{test}} (S_u \cap \Gamma)\right\vert}{|\Gamma|} 
\end{equation}
where $S_u$ is the list generated for user $u$.

This function is related to the \textit{Aggregate Diversity} metric of \cite{adomavicius2012improving} but it looks only at the long-tail part of the item catalog.

\textbf{Cumulative LCR (CLCR):} Since we are modeling the problem of long tail recommendation in a temporal way, we also want to see how the \textit{LCR} changes after each epoch. Note that the cumulative measure is not the cumulative sum of different \textit{LCR} values but it is calculated after the end of each epoch $T_i$ using the entire set of recommendations generated up to that time epoch (i.e. from $T_0$ to $T_i$. More formally:

\begin{equation}\label{eq:aclt}
\begin{split}
    CLCR@T_i=\sum_{t=T_0}^{T_i} \frac{\left\vert \bigcup\limits_{u \in U^t_{test}} (S_u \cap \Gamma)\right\vert}{|\Gamma|} 
\end{split}
\end{equation}

In addition to the aforementioned long-tail diversity metrics, we also evaluate the accuracy of the ranking algorithms in order to examine the diversity-accuracy trade-offs. For this purpose we use the standard \textit{Normalized Discounted Cumulative Gain} (\textit{NDCG}) measure of ranking accuracy. 

The $\lambda$ parameter in Equation 4 has been chosen experimentally for each algorithm and each dataset and the best value selected. For the Epinions dataset, the $\lambda$ values for \textit{Reg}, \textit{Binary}, \textit{Smooth}, \textit{Time Binary} and \textit{Time Smooth} are 0.05, 0.1, 0.0001, 0.0006 and 0.0002, respectively. For the MovieLens dataset, these values are: 0.05, 0.1, 0.1, 0.1 and 0.05, respectively. 

\begin{figure*}[t]
    \centering
    \includegraphics[height=2.9in, width=4.5in]{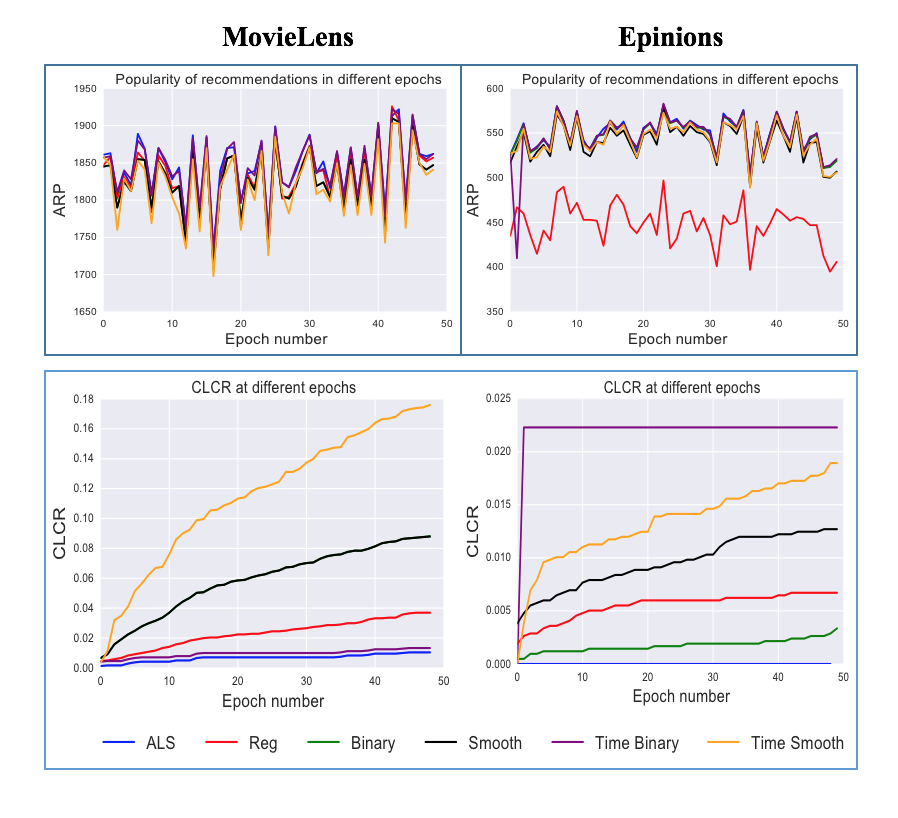}
    \vspace{-0.5cm}
    \caption{The epoch-wise ARP and cumulative LCR (CLCR)}
    \label{fig:ml-results}
\end{figure*}

\begin{table*}[t]
    \centering
    \caption{\label{tab:results}Experimental results. Values not significantly different ($p > 0.05$) from ALS are in italics. Bold values are the best results and are significant improvements over the next best algorithm. * indicates a result significantly worse than the baseline.}
    \begin{tabular}{ l c c c c c c}
        \toprule
        & \multicolumn{3}{c}{MovieLens}  & \multicolumn{3}{c}{Epinions} \\
        \cmidrule(lr){2-4}\cmidrule(lr){5-7}
            & {Average LCR} & {Average NDCG@10} & {Average ARP} & {Average LCR} & {Average NDCG@10} & {Average ARP} \\
            \midrule
        ALS  &    
                0.00059     &  0.262         &  1844 &  0.00000     & 0.0299         &  549         \\
                    \bottomrule
        Reg      &  0.00306 &  \textit{0.261}      &  \textit{1831} & 0.00171     &  0.0243* &  \textbf{447}  \\
          \bottomrule
       Binary &      
       0.00800  & \textit{0.260}    &  1827 & 0.00047    &  \textit{0.0299} & \textit{548}   \\
    \bottomrule
     Smooth     
     &  0.00800 &  \textit{0.259}   &  1827 &0.00309     &  \textit{0.0295}    &  540 \\
    \bottomrule
    
    Time Binary   
    &  \textit{0.00060} &  \textit{0.262}   &  \textit{1843} & \textit{0.00043}    & \textit{0.0299}    & \textit{546}  \\
    \bottomrule
    Time Smooth  
    &  \textbf{0.01530}  &  \textit{0.260}  &  \textbf{1820} &\textbf{0.00344}   & \textit{0.0298}   &  542     \\
    \bottomrule
    \end{tabular}

\end{table*}

\section{Results}
%In this work, we were interested in how a recommender system can compensate for previous omissions and therefore we have not reported the effect of $\lambda$ on each algorithm.  

Figure~\ref{fig:ml-results} shows how long-tail coverage changed over the course of the experiment. The top (\textit{ARP}) figures show the average recommendation popularity in each epoch. The bottom (\textit{CLCR}) figures show the total long-tail coverage considered from the initial epoch to the present one. In the MovieLens dataset, we see that, while the \textit{ARP} score is fairly similar across algorithms, \textit{Time Smooth} xQuAD far out-paces the others in covering more unique items -- it is, of course, specifically designed to incorporate new items at each epoch. 

For Epinions, the plots are quite different. The first notable feature is how good (low) the \textit{ARP} results are for the \textit{Reg} algorithm. However, the \textit{CLCR} results reveal how misleading this metric is. Other algorithms cover many more of the long-tail items. 

Another interesting result is how the \textit{CLCR} for the \textit{Time Binary} algorithm jumps up at first epoch and then remains relatively stable. The reason, as we discovered, is that there are not enough high-quality long-tail items (items with average rating above 3) in this dataset and, therefore, each original ranked list of 100 items for every user contains very few of them. Once these items have been chosen in first epoch, they are not picked again by the algorithm due to its binary nature. The \textit{Time Smooth} algorithm incorporates long-tail items more slowly because they are not rated as highly and therefore not scored highly by the base recommendation algorithm.

Additional results are shown in Table~\ref{tab:results}. On the MovieLens dataset, all algorithms have very similar NDCG, meaning they produce lists with similar ranking quality. There is therefore no cost associated with the improved \textit{CLCR} results shown in Figure~\ref{fig:ml-results}. ARP values are also similar. \textit{LCR} values, however, are quite different for these algorithms, with \textit{Time Smooth} xQuAD incorporating by far the most long-tail items in each list. 
%confirming our argument in the motivating example. The \textit{Time-Smooth} xQuAD has the best \textit{LCR} in almost every epoch followed by the Smooth \textit{xQuAD}. The \textit{Time-Binary} xQuAD and \textit{Binary xQuAd} have very similar performance to the original Rank \textit{ALS}: they do not improve the long-tail recommendation much. 

%The NDCG values are also included in Table~\ref{tab:results} and show that there is little loss (less than 1\%) in ranking accuracy by the introduction of long-tail items. 

The table also includes results for the Epinions dataset. On this dataset, the algorithms behave differently due to the extreme long-tail. For example, the \textit{Reg} algorithm has the best overall \textit{ARP} as Figure~\ref{fig:ml-results} would suggest. However, its \textit{LCR} is worse than \textit{Smooth} xQuAD and \textit{Time Smooth} xQuAD. It is concentrating its long-tail promotion on a small number of long-tail items, and not covering as much of the catalog. It is also not adding many items to recommendation lists, so its impact on overall recommendation outcomes is limited. The NDCG results for Epinions also show no statistically-significant loss for any of the re-ranking algorithms. 

\section{Related Work}

Recommending serendipitous items from the long tail is generally considered to be a key function of recommendation \cite{anderson2006long}, as these are items that users are less likely to know about. Long-tail items are also important for a fuller understanding of users' preferences. Systems that use active learning to explore each user's profile will typically need to present more long tail items. These are the ones that the user is less likely to know about, and where users' preferences are more likely to be diverse~\cite{resnick2013bursting}. 

Item popularity and its impact on recommendation quality has been explored by some researchers \cite{brynjolfsson2006niches,longtailrecsys}. These authors tried to improve the performance of the recommender system in terms of accuracy and precision, given the long-tail in the ratings. Our work, instead, focuses on reducing popularity bias and balancing the representation of items across the popularity distribution.  

A regularization-based approach to improving long tail recommendations is found in \cite{abdollahpouri2017controlling}. One limitation with that work is that it is restricted to factorization models where the long-tail preference can be encoded in terms of the latent factors. This algorithm does not account for differential user tolerance towards long-tail items. A re-ranking approach can be applied to any algorithm, and in our implementation, we also take personalization of long-tail promotion into account. 

There is substantial research in recommendation diversity, where the goal is to avoid recommending too many similar items~\cite{zhou2010solving,castells2011novelty,zhang2008avoiding}, including some research on personalized diversity where the amount of diversification is dependent on the user's tolerance~\cite{eskandanian2017clustering,wasilewski2018intent}. Another similar work to ours is \cite{vargas2012explicit} where authors used a modified version of xQuAD for intent-oriented diversification of search results and recommendations. Another work that also used xQuAD in recommendation is \cite{liu2018personalizing} where the authors used it to improve recommendation fairness in a microlending scenario.

Our work is different from these previous diversification approaches in that it is not dependent on the characteristics of items, but rather on the relative popularity of items. In addition, our work takes the performance of the recommender system at previous times into account in order to compensate for previous omissions. 
Another relevant work to ours is \cite{flairs2019} where authors used xQuAD for long-tail recommendation. However, in that work the long-tail compensation is considered as a one-shot action.

Temporal diversity and novelty has been also explored in \cite{lathia2010temporal} where authors investigated how different algorithms perform in terms of diversity of the recommended item lists over time. Our work is one approach to improve the temporal novelty of the recommendations although our focus is more on the coverage of the item catalog rather than differences across lists.  
\section{Conclusion and Future work}
Many recommendation algorithms have the problem of popularity bias: a few items are recommended frequently while the majority of the items do not appear at all. Research on popularity bias has concentrated on improving individual recommendation lists for each user without taking into account what has been recommended to others. In this work, we showed that a list-wise approach is not effective for long-tail promotion. We presented a temporal approach based on the xQuAD diversification algorithm from information retrieval to improve long-tail recommendation over time. 

Experimental results showed that our approach is capable of recommending more unique long-tail items than the other baselines while maintaining comparable ranking accuracy. In future work, we intend to investigate the effect of epoch size on the performance of the algorithms. In addition, we want to make the algorithm more dynamic by modifying the $\lambda$ parameter in each epoch, depending on performance. Given the differences in algorithm performance across these datasets, we are also interested in developing and fitting parameterized models of long-tail distributions based on the characteristics of the data to go beyond the simple short-head / long-tail division employed here.

\bibliographystyle{ACM-Reference-Format}
\bibliography{main.bib}

\end{document}